\documentclass[superscriptaddress,prb,aps,twocolumn]{revtex4}
\usepackage{graphicx}
\usepackage{subfigure}
\usepackage{bm}
\usepackage{amsmath}
\begin{document}
\title{On Breaking Time Reversal in a Simple, Smooth, Chaotic System}
\author{Steven Tomsovic$^1$, Denis Ullmo$^2$, and Tatsuro Nagano}
\address{Department of Physics, Washington State University,
Pullman, WA 99164-2814}
\address{Laboratoire~de~Physique~Th\'eorique~et~Mod\`eles
Statistiques~(LPTMS),~91405~Orsay~Cedex,~France\\ }
\date{\today}

\begin{abstract}
Within random matrix theory, the statistics of the eigensolutions depend
fundamentally on the presence (or absence) of time reversal symmetry.
Accepting the Bohigas-Giannoni-Schmit conjecture, this statement extends
to quantum systems with chaotic classical analogs.  For practical reasons,
much of the supporting numerical studies of symmetry breaking have been
done with billiards or maps, and little with simple, smooth systems.
There are two main difficulties in attempting to break time reversal
invariance in a continuous time system with a smooth potential.  The
first is avoiding false time reversal breaking.  The second is locating a
parameter regime in which the symmetry breaking is strong enough to
transition the fluctuation properties fully toward the broken symmetry
case, and yet remain weak enough so as not to regularize the dynamics
sufficiently that the system is no longer chaotic.  We give an example of
a system of two coupled quartic oscillators whose energy level
statistics closely match those of the Gaussian unitary ensemble, and
which possesses only a minor proportion of regular motion in its phase
space.
\end{abstract}

\pacs{05.45.Mt, 03.65.Sq}
\maketitle

Since its introduction into nuclear physics by Wigner~\cite{wigner},
random matrix theory (RMT) has grown to encompass a broad variety of
applications, and can be viewed as a significant portion of the foundation
of the statistical mechanics of finite systems.  One of the central
tenets of RMT emphasized by Wigner was that the ensembles carry no
information other than that required by the symmetries of the
system.  A very important symmetry is time reversal which determines
whether the ensemble is composed of real symmetric matrices, Gaussian
orthogonal ensemble (GOE - good symmetry), or complex hermitian matrices,
Gaussian unitary ensemble (GUE - broken symmetry).  This distinction
leads to quite different predictions for the statistical properties of
both the energy levels and the eigenfunctions.  In fact, Wigner proposed
it as a test of time reversal invariance in the strong interaction using
slow neutron resonance data~\cite{tri}.  His suggestion was not fully
realized until more than twenty years later~\cite{fkpt}.

Time reversal is well known to be an antiunitary symmetry in quantum
mechanics, and Robnik and Berry generalized the criterium for expecting
the statistical properties of the GOE to include any antiunitary
symmetry~\cite{rb}.  An elementary example of an antiunitary symmetry
would be the product of some reflection symmetry and time reversal.  Some
of the early investigations of noninvariant systems were of Aharanov-Bohm
chaotic billiards~\cite{br}, symmetry breaking quantum
maps~\cite{izrailev}), and combinations of magnetic and scalar
forces~\cite{selig}.

\begin{figure}
\includegraphics*[width=3.5in,height=5.5in]{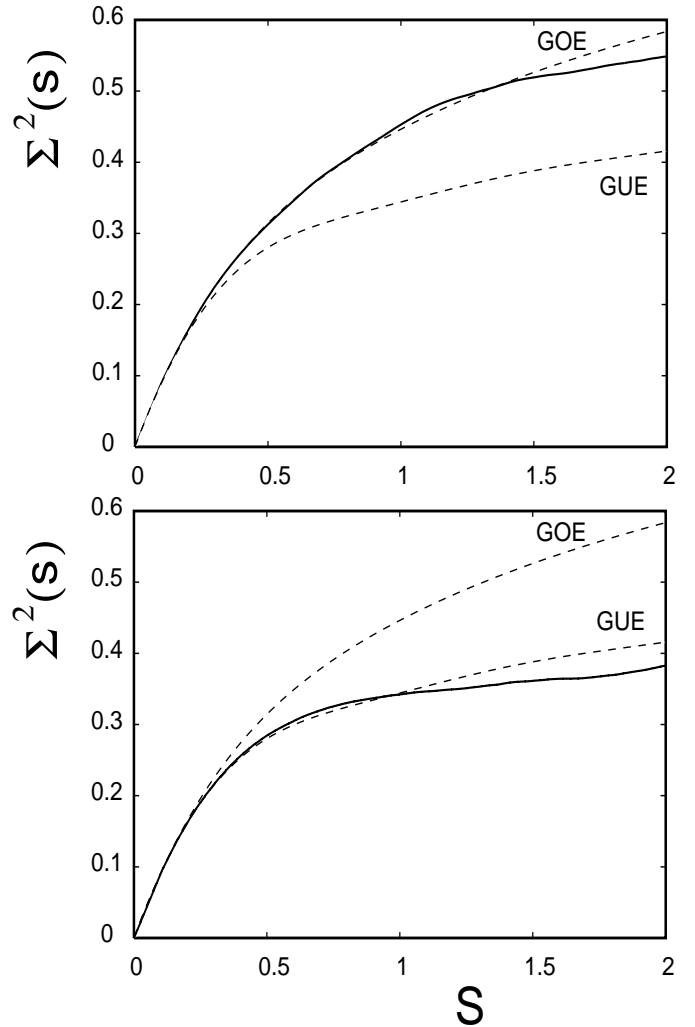}
\caption{The number variance for the quartic oscillators (solid lines)
compared to the GOE and GUE (dashed lines).  The upper panel shows the
statistics for the hidden symmetry case, i.e. with the Hamiltonian term
in Eq.~(4)) with
$(\epsilon = 0.5, \lambda = -0.65)$.  The lower panel shows the results
using the Hamiltonian term of Eq.~(11) and $(\epsilon = 1.0, \lambda =
-0.80)$.  Spectra containing a total of $600$ levels were used to
generate the statistics.  The lowest $50$ eigenvalues were dropped.}
\label{stat}
\end{figure}

There is no known mechanical-type system of a particle moving under the
influence of a simple, closed, smooth potential whose dynamics is
rigorously proven to be fully chaotic, independently of the question of
anti-unitary symmetry; we exclude diffusive dynamics associated with
random potentials.  For some period of time, the $x^2y^2$ quartic
potential was a prime candidate, but eventually stable trajectories were
found~\cite{swedes}.  Nevertheless, there exists a family of quartic
potentials with $x^2y^2$ as a limiting case whose fluctuations closely
approximate GOE statistics, and whose classical dynamics contain
negligible phase space zones of stable trajectories.  Therefore,to
construct a close approximation of a Hamiltonian without an anti-unitary
symmetry and GUE statistics, we begin by considering the symmetry
preseverving two-degree-of-freedom coupled quartic oscillators whose
Schr\"odinger equation is given by
\begin{equation}
\label{schro}
\hat H_0 \Psi(\vec r) = -{\hbar^2\nabla^2 \over 2m} \Psi(\vec r) + \hat
V(\vec r)
\Psi(\vec r) = E\Psi(\vec r)
\end{equation}
The potential can be expressed as
\begin{equation}
\label{pot}
\hat V(\vec r) = a(\lambda)\left[{x^4 \over b} + b y^4  +2\lambda x^2 y^2
\right]
\end{equation}
where $a(\lambda)$ is a convenient constant, $b\ne 1$ lowers the
symmetry, and $\lambda$ gives the strength of the coupling; $(b=1,
\lambda =-1)$ is equivalent to $V(\vec r) = x^2y^2$ by a $\pi/4$
rotation of the $(x,y)$ coordinates.  The corresponding classical
Hamiltonian is
\begin{equation}
\label{classham}
H_0(\vec r,\vec p) = {\vec p^2 \over 2m} + V(\vec r)
\end{equation}
This system is symmetric under time reversal and reflections with
respect to $x$ and $y$.  For strong couplings, $\lambda \le -0.6$, the
statistics have been shown to agree extremely well with the GOE
predictions~\cite{btu}.  As a first attempt to break time reversal
symmetry, we add the following term to the quantum Hamiltonian ($\hat H =
\hat H_0 + \hat H_1$),
\begin{equation}
\label{pert}
\hat H_1 = {i\hbar\epsilon \over 2m} \left( r{\partial \over \partial r} r
+ {\partial \over \partial r} r^2 \right)
\end{equation}
where $r$ is the radial polar variable.  $\hat H_1$ breaks time reversal
invariance in the Hamiltonian without altering the original reflection
symmetries, and thus does not admit an antiunitary symmetry from any
combination of reflection and time reversal.  This term was chosen to
maintain a scaling property of the eigensolutions of the quartic
oscillators due to the homogeneity of the potential~\cite{btu}.  However,
$\hat H_1$ gives us an excellent example of false symmetry breaking. 
In fact, it turns out that $\hat H_1$ is given
by the cross terms arising from a vector potential ${\bf A}(\vec r) =
\epsilon r^2 \hat r$ (which can also be expressed as the gradient of a
scalar function); i.e.
\begin{equation}
\label{vecpot}
\hat H_1 = {i\hbar \over 2m} \left( \nabla \cdot {\bf A}(\vec
r) +  {\bf A}(\vec r) \cdot \nabla \right)
\end{equation}
This  means  that, up  to the  addition  to  the  potential of  a  term
$\epsilon^2 r^4$, $H_1$ can be  understood as deriving from a magnetic
field ${\bf B} = \nabla \times  {\bf A} = 0$, which obviously will not
change any  physical quantity.  And indeed,  it is possible  to make a
gauge  transformation and rewrite  the wave  function using  the Dirac
substitution
\begin{equation}
\label{gauge}
\Psi(\vec r) = \exp \left({i\over \hbar}\int^{\vec r} {\bf A}(\vec
r^\prime) \cdot {\rm d}\vec r^\prime \right) \Psi^\prime(\vec r) = {\rm
e}^{i r^3 \over 3\hbar}  \Psi^\prime(\vec r)
\end{equation}
in  order to  cancel $\hat  H_1 $ plus the aforementioned $\epsilon^2 r^4$
term.   Furthermore, the transformation is single valued since ${\bf B} =
0$ implies
\begin{equation}
\label{sv}
\oint {\bf A}(\vec r^\prime) \cdot {\rm d}\vec r^\prime = 0 \; .
\end{equation}
The  expectation,   following  the  considerations   of~\cite{rb},  is
therefore to  find GOE statistics and  not those of the  GUE.  See the
upper panel  of Fig.~(\ref{stat}), which compares  the number variance
of the system  (i.e. the variance of the number of  levels found in an
energy  interval  of width  $s$  scaled  locally  to mean  unit  level
density) with the predictions of the GOE and GUE.
The GOE statistics are closely matched to a mean level spacing and a bit
beyond.  The parameter $\epsilon$ was chosen slightly greater than unity
because this forced the eigenlevels through a couple of avoided
crossings, which would have been sufficient to push the spectral
fluctuations toward GUE if the symmetry was not being falsely broken.

 From a classical perspective, the Hamiltonian, not incorporating the jauge
transformation, is
\begin{equation}
\label{classham2}
H(\vec r,\vec p) = {p_r^2 - 2\epsilon r^2p_r \over 2m} + {p_\theta^2 \over
2mr^2} + V(r,\theta)
\end{equation}
which itself appears to violate time reversal symmetry, as well as in
Hamilton's equations of motion:
\begin{eqnarray}
\label{heq}
\dot r &=& {1\over m}(p_r-\epsilon r^2)\hskip 4em
\dot p_r = - {\partial V(r,\theta) \over \partial r} + 2{\epsilon \over
m} r p_r \nonumber \\
\dot \theta &=& {p_\theta \over mr^2} \hskip 7.25em
\dot p_\theta = -{\partial V(r,\theta) \over \partial \theta}\nonumber \\
\end{eqnarray}
However even without considering a canonical transformation, just by
converting to the Lagrangian description of the dynamics using the left
hand side equations, it turns out that
\begin{equation}
\label{lagrange}
{\cal L} = {1\over 2}(m \dot r^2 + mr^2 \dot \theta^2) - V(r,\theta)
+ {\epsilon^2 \over 2m} r^4 + \epsilon r^2 \dot r
\end{equation}
The final term, which appears to break the symmetry, cannot enter the
equations of motion.  They are invariant under addition of any total
time derivative.  It is not always obvious, a priori, whether a symmetry
breaking term leads to false symmetry breaking or not.

If we multiply the symmetry breaking term by any function of $\theta$, it
can no longer be a total time derivative, nor can the vector potential
be expressed as the gradient of a scalar function.  Consider
\begin{equation}
\label{pert2}
\hat H_1^\prime = {i\hbar\epsilon \over 2m} \cos^2 \theta \left(
r{\partial \over\partial r} r + {\partial \over \partial r} r^2 \right)
\end{equation}
The  vector  potential  becomes   ${\bf  A}(\vec  r)  =  \epsilon  r^2
\cos^2\theta \hat r$,  and therefore ${\bf B} =  2 \epsilon xy/r \neq
0$.  The  integrals $\int^{\vec r}  {\bf A}(\vec r^\prime)  \cdot {\rm
d}\vec r^\prime$  and $\oint {\bf A}(\vec r^\prime)  \cdot {\rm d}\vec
r^\prime$ are  path dependent.  The Dirac substitution  is not useful,
and false symmetry breaking is not an issue.

\begin{figure}
\includegraphics*[width=3.5in,height=2.7in]{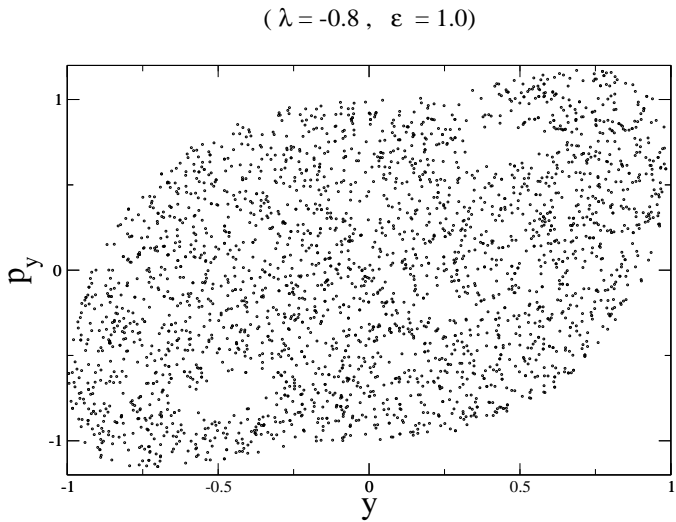}
\caption{The $x=0$  Poincar\'e surface of section  for the Hamiltonian
with the term from Eq.~(11), and parameters ($\lambda = -0.8, \epsilon
= 1$)  as in  the bottom  panel of Fig.~(1).  The coordinates  $y$ and
$p_y$ are in scaled unit.}
\label{sos}
\end{figure}

Driving the eigenvalues at numerically attainable energies through a
couple of avoided crossings forces $\epsilon$ to be chosen in the
neighborhood of unity.  This has a strong regularizing effect on the
nature of the dynamics.  See the surface of section in
Fig.~(\ref{sos}) for the case ($\lambda = -0.8, \epsilon = 1$).
Only its spectrum gives number variance statistics close to the GUE
results; see the lower panel of Fig.~(\ref{stat}).

To summarize, we have given an example of a simple, continuous, dynamical
system that comes close to generating GUE statistics.  It is surprisingly
difficult to find an essentially, fully chaotic system that does so.  The
pitfall of false time reversal breaking can lead to symmetries that are
quite well hidden, and the addition of a vector potential to a dynamical
system has a strong tendency to move the system away from fully developed
chaos.

We gratefully acknowledge support from ONR grant N00014-98-1-0079 and NSF
grant PHY-0098027.

\end{document}